%% file: DSCC2019LearningFinal.tex
\documentclass[twocolumn,10pt]{asme2e}

\usepackage{cite}
\usepackage{amsmath}
\usepackage{graphicx}
\usepackage[font=footnotesize,labelfont=bf,skip=0pt]{caption}
\usepackage{epstopdf}
\usepackage{amssymb}
\usepackage{enumerate}
\usepackage{etaremune}
\usepackage{setspace}
\usepackage{comment}
\usepackage{color}
\usepackage[table]{xcolor}
\usepackage{soul,xcolor}
\setstcolor{red}
\usepackage{wrapfig}
\usepackage{lipsum}
\usepackage[colorlinks,
linkcolor=black,
anchorcolor=black,
citecolor=black]{hyperref}
\definecolor{blue}{RGB}{0, 111, 213}
\definecolor{blue0}{RGB}{100, 211, 100}
\definecolor{red}{RGB}{150, 11, 23}
\usepackage{soul}
\usepackage{booktabs}
\usepackage{float}
\usepackage{mathrsfs}
\DeclareMathAlphabet\EuScript{U}{eus}{m}{n}
\SetMathAlphabet\EuScript{bold}{U}{eus}{b}{n}
\usepackage{pgfplots} 
\pgfplotsset{compat=newest} 
\pgfplotsset{plot coordinates/math parser=false} 
\newlength\figureheight 
\newlength\figurewidth 
\usetikzlibrary{mindmap}
\usepackage{placeins}
\usetikzlibrary{arrows}
\usepackage[mathscr]{eucal}
\usepackage{tikz-3dplot} 
\usepackage{pgfplots}
\pgfplotsset{compat=newest}
\usepackage{tikz}
\tikzset{
	MyPersp/.style={scale=1.8,x={(-0.8cm,-0.4cm)},y={(0.8cm,-0.4cm)},
		z={(0cm,1cm)}},
	MyPoints/.style={fill=white,draw=black,thick}
}
\usepackage{graphicx}
\usepackage[table]{xcolor}
\usepackage{pgfgantt}
\ganttset{group/.append style={orange},
	milestone/.append style={red},
	progress label node anchor/.append style={text=red}}
\usepackage{array, booktabs}
\usepackage{graphicx}
\usepackage{caption}
\usepackage[algo2e]{algorithm2e} 
\usepackage{algorithm} 
\usepackage{algpseudocode}

\confshortname{DSCC 2019}
\conffullname{the ASME 2019 Dynamic Systems and Control Conference}

\confdate{8-11}
\confmonth{October}
\confyear{2019}
\confcity{Park City, Utah}
\confcountry{USA}

\papernum{DSCC2019-9186}

\title{A Preliminary Study on A Physical Model Oriented Learning Algorithm with Application to UAVs}

\author{Minghui Zheng\thanks{Address all correspondence to this author.}
	\affiliation{
		Department of Mechanical \\ and Aerospace Engineering\\
		University at Buffalo\\
		Buffalo, New York 14260\\
		Email: mhzheng@buffalo.edu
	}	
}

\author{Zhu Chen
	\affiliation{
		Department of Mechanical \\ and Aerospace Engineering\\
		University at Buffalo\\
		Buffalo, New York 14260\\
		Email: zhuchen@buffalo.edu
	}
}

\author{Xiao Liang
	\affiliation{
		Department of Civil, Structural \\and Environmental Engineering \\
		University at Buffalo\\
		Buffalo, New York 14260\\
		Email: liangx@buffalo.edu
	}
}

\begin{document}
\maketitle
\thispagestyle{empty}
\pagestyle{empty}
\begin{abstract} 
This paper provides a preliminary study for an efficient learning algorithm by reasoning the error from first principle physics to generate learning signals in near real time. Motivated by iterative learning control (ILC), this learning algorithm is applied to the feedforward control loop of the unmanned aerial vehicles (UAVs), enabling the learning from errors made by other UAVs with different dynamics or flying in different scenarios. This learning framework improves the data utilization efficiency and learning reliability via analytically incorporating the physical model mapping, and enhances the flexibility of the model-based methodology with equipping it with the self-learning capability. Numerical studies are performed to validate the proposed learning algorithm.
\end{abstract}

\section{Introduction}
Unmanned aerial vehicles (UAVs) have achieved great progress and been applied to many areas including traffic monitoring \cite{kanistras2015survey,li2018unmanned}, structural health monitoring \cite{Liang2018-1,Liang2018-2,Liang2018-3,Liang2018-4}, and search and rescue \cite{ahmadzadeh2006cooperative,doherty2007uav,goodrich2008supporting}. UAVs are expected to have autonomy and maneuverability in these applications. Targeting these properties along with explosively increased computational power, attention has been directed to data-driven techniques such as machine learning and reinforcement learning that enable UAVs to execute tasks with full autonomy based on prior training. While these techniques greatly increase the flexibility and responsiveness of UAVs, the training process usually takes time and may not be able to perform in near real time. This has severely limited UAV's application in time-intensive and data-scarcity situations with an emergency in which extensive prior training is usually impossible. 

An increasing number of techniques have been explored and developed for UAVs towards greater autonomy, accuracy, and reliability. While this area covers rather broad topics including perception, image processing, decision making, learning, planning, navigation, identification, and control, herein we focus on a brief review on the techniques that particularly aim to acquire planning and control strategy for the UAVs to execute certain tasks with aggressive maneuvers. These techniques, among others, include reinforcement learning, imitation learning, planning and tracking control, disturbance observer, and iterative learning control. The above-mentioned techniques generally fall into two categories: the data-driven learning and the physical model-based control.

The data-driven learning techniques play critical roles in presence of large uncertainties. The imitation learning approach is within the supervised learning methodology, in which the training data that includes the task and the expert's actions is provided to enable the UAV to best mimic the expert's action based on a large learning data set  (e.g., \cite{hussein2017imitation}). 
The reinforcement learning techniques, such as value function approach, policy gradient method, and probabilistic direct policy learning have also been developed and applied to UAV-related areas (e.g., \cite{5611206,zhang2016learning}). It generally aims to map the states to control actions and maximize the value functions based on certain defined reward functions. There are also other learning techniques such as inverse reinforcement learning  (e.g., \cite{choi2017inverse}) and deep reinforcement learning (e.g., \cite{challita2018cellular}). It is worth mentioning that there exists the model-based reinforcement learning (e.g., \cite{polydoros2017survey}) which we still classify as the data-driven instead of the model-based methodology, as the latter one in this paper particularly refers to the physical model oriented techniques. These above-mentioned data-driven learning algorithms can be an alternative to existing control techniques when the dynamic models are well known and have shown great flexibility when the dynamic models are not well established or even not available. Although the increasing computation power has greatly boosted the data-driven learning techniques, many of them are still restricted to small academic applications or highly dependent on extensively trials (as discussed in \cite{schaal2010learning}). In an emergency, it is challenging for these data-driven learning algorithms to immediately handle unexpected situations or carry out certain tasks with limited training.

The model-based control guarantees the UAV's tracking performance in terms of tracking accuracy and robustness to disturbance. These control techniques mainly fall into two categories: the feedback and the feedforward control. Among others, the trajectory generation and tracking algorithms enable UAVs to accurately and aggressively maneuver in a cluttered and indoor environment with reduced uncertainties (e.g., \cite{mellinger2011minimum}). The disturbance observer for UAVs estimates and compensates large unknown external disturbances exerted on UAVs and enhances the robustness of UAV's attitude control (e.g., \cite{zheng2017UAVDOB,mishra2017disturbance}). The feedforward techniques provide additional flexibility to further enhance the UAV's tracking performance. These techniques include adaptive feedforward control (e.g., \cite{gruning2012feedforward}) and ILC (e.g., \cite{schoellig2012optimization,hehn2013iterative}). Particularly, ILC is developed for iterative performance enhancement in structured and repetitive tasks(\cite{zheng2017design,zheng2018systematic,wang2018robust,schoellig2012optimization}). These above-mentioned model-based control techniques play important roles in maintaining the UAV's stability and robustness to modeling uncertainties, some design trade-offs including the fundamental water-bed effect in feedback systems severely limits the flexibility of these techniques for fast-changing tracking scenarios.

Most of the current autonomy solutions for UAVs are derived from the data-driven methodology. Although these techniques have significantly increased the flexibility and capacities of UAVs, there still exist underlying challenges such as high dependence on extensive training and high computational cost on the way towards full autonomy. On the other hand, the control technique is a model-based methodology that has been well developed to guarantee the UAV's stability and tracking accuracy with yet limited autonomy and flexibility. This paper targets a new way to establish a rapprochement between data-driven and physical model oriented methodologies and aims to enable the UAVs to (1) learn from historical experience of itself or others with sufficient efficiency, robustness, and self-awareness of the physical constraints, and (2) fly with high precision and maneuverability in a cluttered and unstructured environment with limited prior training. The proposed algorithm is a physical model oriented learning process that can be performed in near real time for autonomous systems to achieve greater learning efficiency and robustness.

Our previous work in \cite{liang2018scalable} introduced a model-based learning among the UAVs that share the same dynamics, and performed a comparison study with a reinforcement learning algorithm. In this paper, we generalize the learning algorithm by enabling the learning from UAVs with different dynamics and flying in different scenarios. The remaining of the paper is organized as follows. Section II briefly introduces the model-based learning algorithm. Section III and Section IV generalize the learning algorithm into heterogeneous scenarios and agents with numerical verifications. Section V unifies the two extensions in one framework and provides step-by-step implementation details as well as corresponding numerical verifications. Section VI concludes this paper with some future work discussion.

\section{Learning Algorithm Overview}

\begin{figure}[!htbp]
	\begin{center}
		\includegraphics[width=0.45\textwidth]{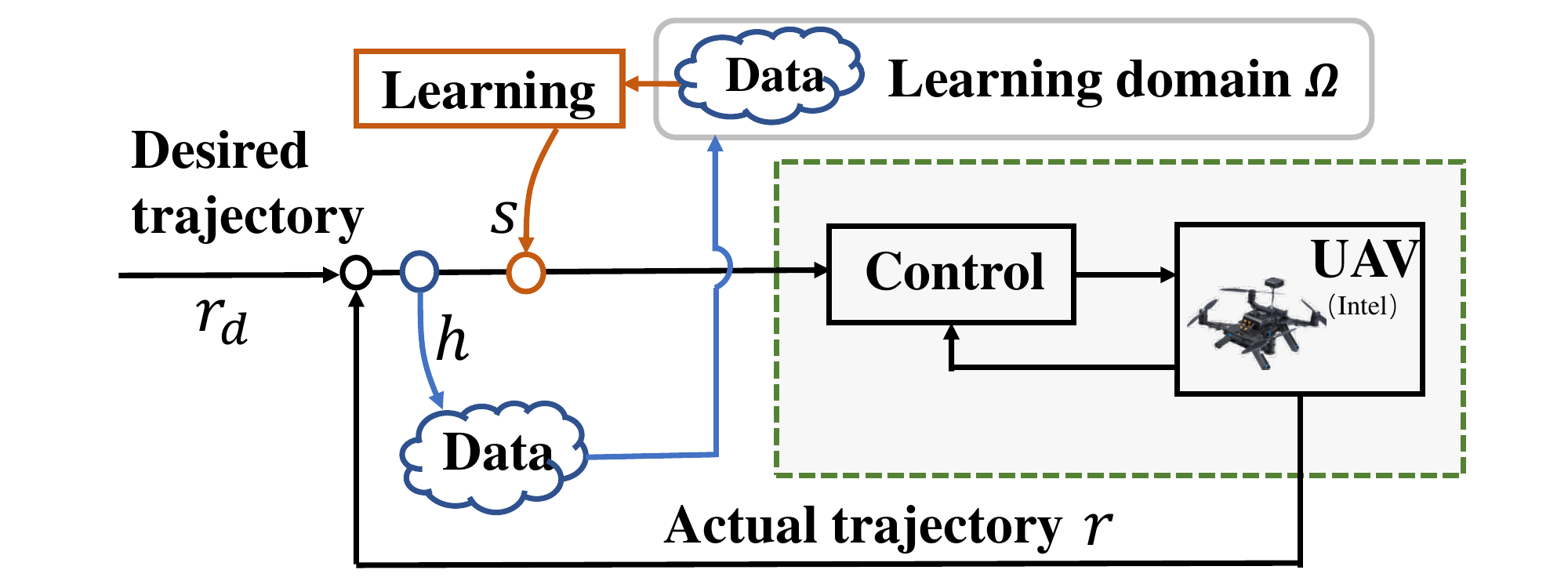}
		\caption{Proposed learning in the target UAV's loop.}
		\label{fig:Control}
	\end{center}
\end{figure}

The proposed framework consists of two sets of UAVs (the target UAVs and the training UAVs). For a system $\EuScript{G}$, we use $\EuScript{G}(g)$ to denote system $\EuScript{G}$ with system's variable $g$, and $\EuScript{G}\{q\}$ to denote the output of system $\EuScript{G}$ with the input signal $q$. The learning domain for the target UAV is denoted as $\Omega{=}\{k_0,..., k,..., k_f\}$, which means the target UAV is learning from multiple training UAVs that are tagged from $k_0$ to $k_f$. As illustrated in Fig.~\ref{fig:Control}, the target UAV follows the planned trajectory $r_d$ while receiving the learning signal $s$ that is generated in near real time from the training UAVs. Meanwhile, the performance criterion $h$ (i.e., tracking error) is returned to the data set which will be utilized to optimize the target UAV's performance and learning efficiency. The learning algorithm exploits the design strategy for the learning signal $s$ that is generated from the UAVs in $\Omega$ to enhance the tracking performance of the target UAV. 

Denote $T$ and $F$ as the dynamic systems from $r_d$ to $h$ and from $s$ to $h$, respectively, we have
\begin{equation}\label{sys}
h=T\{r_d\}+F\{s\}
\end{equation}
We only consider the position control loop of the UAVs and ignore the attitude control loop whose bandwidth is much higher, the dynamics, $T$ and $F$, are reasonably assumed as linear time-invariant (LTI) systems (\cite{liang2018scalable,mellinger2011minimum}).

The learning signal $s$ will be generated from the flying data of UAVs in $\Omega$ which includes the learning signals $s_{k}$'s and the tracking error signals $h_{k}$'s. Such a learning data file ($s_k,h_k,~k{\in}\Omega$) goes through the robust filters ($\alpha_{k}$) and the learning filters ($L_k$) which are designed in the next section. We propose the following learning algorithm
\begin{equation}\label{sj}
s = \sum_{k\in \Omega} (\alpha_{k}s_{k} + L_{k}\{h_{k}\})
\end{equation}
Then the dynamic relationship from the training UAVs flying in the same scenarios to the target UAV can be established as follows
\begin{equation}
\begin{split}
h&=F\{\sum_{k\in \textcolor{black}{\Omega}} (\alpha_{k}s_{k}+ L_{k}\{h_{k}\})\}+T\{r_d\}\\
&=\sum_{k\in \textcolor{black}{\Omega}}\left[\alpha_{k}F\{s_k\}+(FL_k)\{h_k\}\right]+T\{r_d\}
\end{split}
\end{equation}
Considering that $F\{s_k\}=h_k-T\{r_d\}$,
we have
\begin{equation}
\begin{split}
h&=\sum_{k\in \textcolor{black}{\Omega}}\left[\alpha_{k}(h_k-T\{r_d\})+(FL_k)\{h_k\}\right]+T\{r_d\}\\
&=\sum_{k\in \textcolor{black}{\Omega}}\left[(\alpha_{k}I+FL_k)\{h_k\}-\alpha_{k}T\{r_d\}\right]+T\{r_d\}
\end{split}
\end{equation}
where $\alpha_{k}$ and $L_k$ are designed such that $\small{||h|| < \min \{||h_k||{,}~k{\in}\Omega\}}$. One way to design the learning filter is to find a $L_k$ such that
\begin{equation}
\alpha_{k}I+FL_k \approx 0
\end{equation}
which implies 
\begin{equation}
L_k\approx \alpha_{k} F^{-1}
\end{equation}

The above learning algorithm is only applicable for identical UAVs flying in the same scearios, motivated by ILC. In the following two sections, we will generalize the learning algorithm to include respectively transformable heterogeneous scenarios and agents with numerical verifications. These generalizations enable the learning among different UAVs flying in transformable scenarios.

\section{Learning from heterogeneous scenarios}

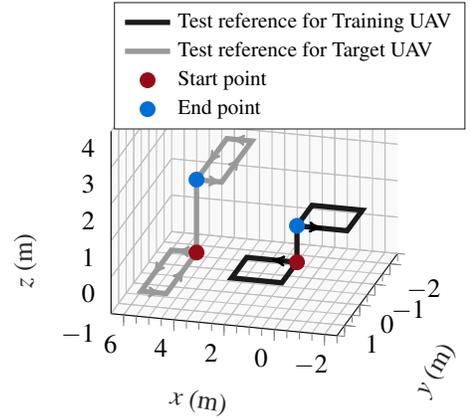
\begin{figure}[!htbp]
	\centering
	\input{figure/Reference2.tex}
	\caption{References for the training UAV and the target UAV.}
	\label{fig:Reference}
\end{figure}

Considering that the flying scenarios of the training and the target UAVs are usually different, we will generalize the learning from identical or scaled scenarios to the scenario that from the heterogeneous ones. The key idea is to establish a transformation relationship between different flying scenarios and to employ it analytically into the learning framework. As a study example, we use the references to represent the flying scenarios. One such example is illustrated in Fig.~\ref{fig:Reference}, in which the target UAV's reference can be mapped to $k^{th}$ training UAV's reference by a matrix. 

\vspace{-15pt}
\vspace{-15pt}

\subsection{Formulation}

For each UAV in the learning domain $\Omega$, we explicitly introduce scaling matrices for different scenarios, i.e., $\Lambda_k \in R^{3\times 3}$ is the scaling matrix for the reference of the $k^{th}$ UAV in $\Omega$. Assume that the $k^{th}$ training UAV's desired trajectory is $\Lambda_k r_d$. We assume that the position control loop of the UAV is decoupled, that is, $T$ and $F$ are diagonal-matrix systems, and $\Lambda_k$'s are diagonal matrices. 

Correspondingly, the learning algorithm is modified as
\begin{equation}\label{eq:learningsignal}
s = \sum_{k\in \Omega} \Lambda_k^{-1} [\alpha_{k}I\{s_{k}\} + L_{k}\{h_{k}\}]
\end{equation}
where $\alpha_k$ and $L_k$ are respectively a single-input single-output system and a three-input three-output system that are to be designed, and $I$ is a three-by-three identity matrix. With this learning signal in (\ref{eq:learningsignal}), for the target UAV, we have
\begin{equation}
h=T\{r_d\}+F\{s\}
\end{equation}
With this modification, the learning dynamics from the UAVs in $\Omega$ to the target UAV can be derived as follows
\begin{equation}\label{eq:error0}
\begin{split}
	h&=T\{r_d\}+ \sum_{k\in \Omega} \Lambda_k^{-1} [(\alpha_{k}F)\{s_{k}\} + (FL_{k})\{h_{k}\}]
\end{split}
\end{equation}
Considering that for each training UAV, we have
\begin{equation}
h_k=T\{\Lambda_kr_d\}+F\{s_k\}
\end{equation}
Therefore,
\begin{equation}\label{eq:error1}
\begin{split}
h&=T\{ r_d\}+ \sum_{k\in \Omega} \Lambda_k^{-1} [\alpha_{k}(h_k-T\{\Lambda_kr_d\}) + (FL_{k})\{h_{k}\}]\\
&=T\{ r_d\}+ \sum_{k\in \Omega} \Lambda_k^{-1} (-\alpha_{k} T\{\Lambda_kr_d\})\\
& ~~~~+  \sum_{k\in \Omega} \Lambda_k^{-1} (FL_{k})\{h_{k}\}+ \sum_{k\in \Omega} \Lambda_k^{-1} \alpha_{k}\{h_k\}\\
&=  \sum_{k\in \Omega} \Lambda_k^{-1} (FL_{k}+\alpha_{k}I)\{h_{k}\}
\end{split}
\end{equation}

\textit{Stability and learning convergence:} 
The learning signal $s$ will be injected to the feedforward loop of the UAV tracking system, as shown in Fig.~\ref{fig:Control}. Therefore, the proposed learning algorithm will not affect the stability of the UAV tracking system. Based on dynamic relationship between the performance criterion $h$ (i.e., tracking error) of the target UAV and the ones ($h_k,~k\in \Omega$) of the training UAVs in Eq.~(\ref{eq:error1}), the convergence of the learning algorithm in Eq.~(\ref{eq:learningsignal}) for heterogeneous scenarios can be obtained as 
\begin{equation}\label{eq:opt}
\| \begin{bmatrix} \Lambda_0^{-1}(\alpha_1 I+FL_1), ..., \Lambda_N^{-1}(\alpha_N I+FL_N) \end{bmatrix}\|_\infty{<}1/N
\end{equation}
The learning filters will be designed such that the above condition is satisfied \cite{zheng2017design}.

\begin{figure}[!htbp]
	\centering 
	\includegraphics[width=0.45\textwidth]{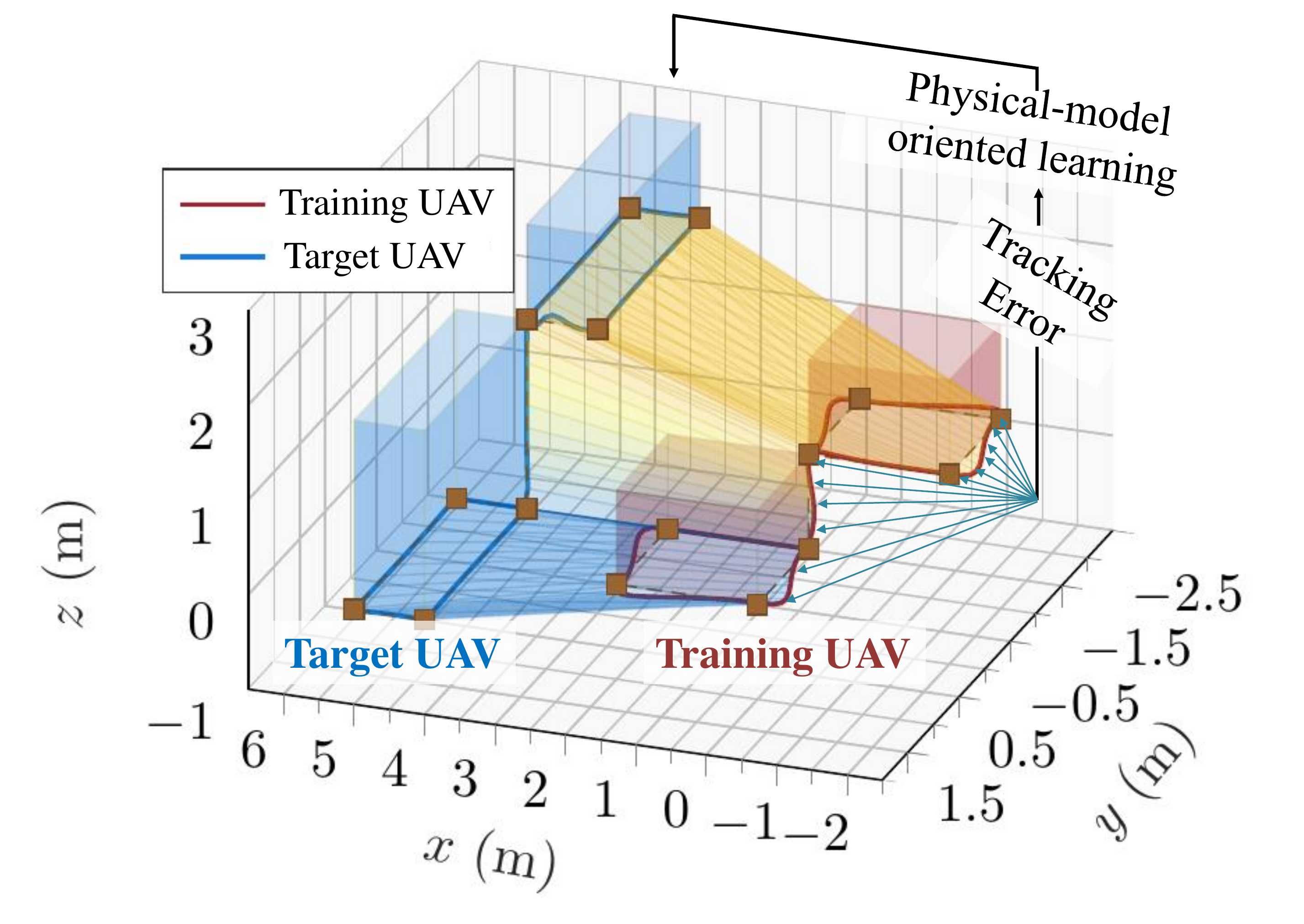} \vspace{5pt}
	\includegraphics[width=0.45\textwidth]{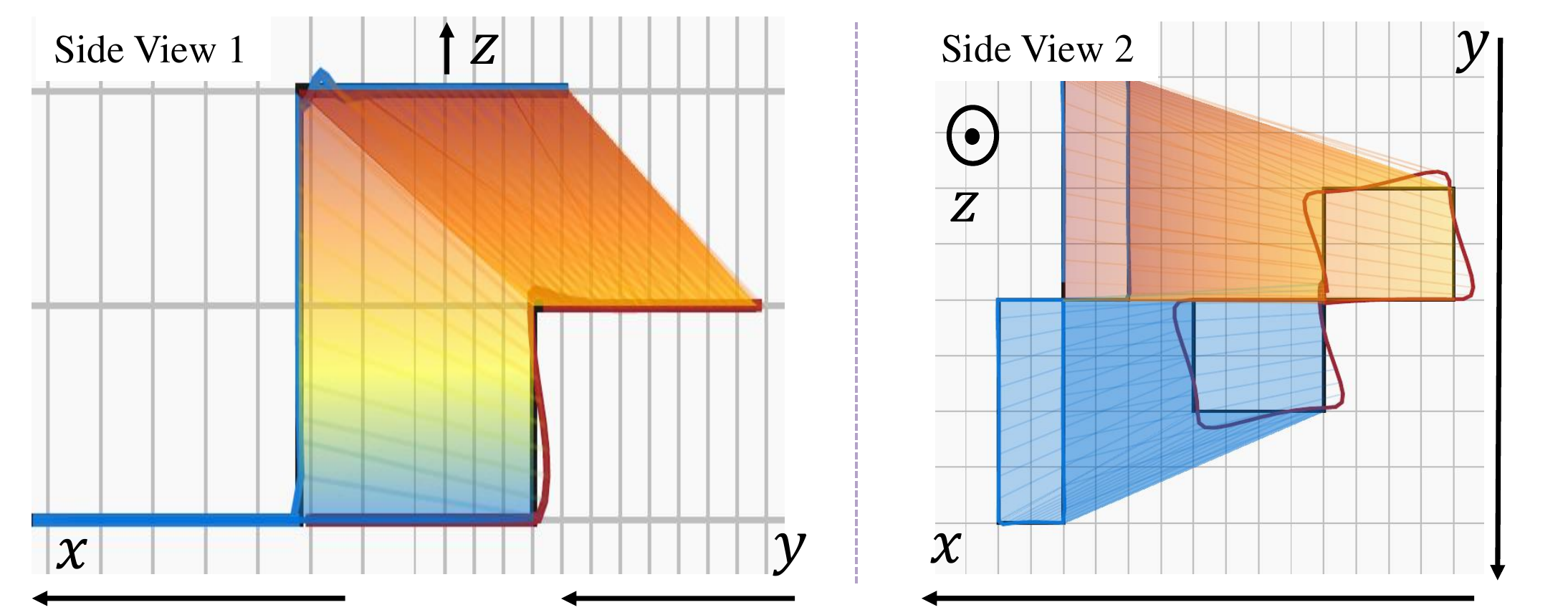}
	\vspace{5pt}
	\caption{Trajectory tracking performance comparison between the training and the target UAVs tracking different references.} \label{fig:Preliminary}
\end{figure}

\subsection{\textcolor{black}{Numerical Verification}}
We provide the numerical verification for the learning from UAVs flying in transformable scenarios. The setting for the numerical study is listed as follows: one target UAV and one training UAV fly respectively with the references of $r_d$ and $r_1$, as illustrated in Fig.~\ref{fig:Reference}. The target UAV learns from the training UAV in which the learning activation can be just one sampling time in advance (near real time). The learning filter is designed such that $\|\Lambda_1^{-1}(\alpha_{1}I+FL_1)\|$ is small. In this case, $\alpha_{k}$ is chosen as 1 and $L_1$ is designed to approximate the inverse of the dynamics $F$. The simulation results are provided in Fig.~\ref{fig:Preliminary}. The top figure illustrates the learning process and the performance enhancement via the proposed learning algorithm: the target UAV is able to track the scaled reference with very small overshoot and oscillations. The target UAV's tracking performance has been significantly enhanced.

\section{Learning from heterogeneous agents}
This section is to further increase the generality and flexibility of the proposed learning algorithm via enabling the target UAV to learn from different UAVs. To do this, we explicitly integrate the dynamics mapping among heterogeneous agents in the learning algorithm to remove the learning signals' dependence on the UAVs' dynamics.

\subsection{Formulation}
In this section, we explicitly involve the UAV's dynamics into the learning algorithm. Considering a quad-rotor UAV flying in a three-dimensional space: the UAV has six degrees of freedom, including the position $r{=}[x,y,z]^T$ and the attitude $\EuScript{\xi}{=}[\phi,\theta,\psi]^T$, in which $(\phi,\theta,\psi)$ denote the roll, pitch, and yaw angles, respectively. Using the differential flatness property and the transformation method from the desired reference ${r}_d$ to the attitude $\xi$ detailed in \cite{mellinger2011minimum}, the desired force and yaw angle can be mapped onto the desired attitude and thrust. Based on this, the control of a quad-robot consists of the inner attitude control loop and the outside position control loop. The attitude controller can track a desired attitude quickly with a very short transient period because the bandwidth of the attitude control loop is usually much higher than that of the position control loop. Therefore, we reasonably ignore the attitude dynamics when designing the learning controller for the UAV's position control. This results in a multi-input multi-output LTI system.

Assume that the open-loop dynamics of the target UAV and the training UAV $k$ could be represented as transfer function matrices $G$ and $G_k$, respectively. We assume that the mapping from $G_k$ to $G$ could be represented by a transfer function matrix $M_k$, i.e., $G=M_kG_k$. 
It is worth noting that, based on the definition of $T$ and $F$, we have
\begin{equation}
T=(1+G)^{-1},~\text{and}~ F=-G(1+G)^{-1}
\end{equation}

Considering that the position controller of the UAV guarantees the transfer function from the reference $r_d$ to the output $r$ to be approximately 1, we have $\|G\| \gg 1$ and
\begin{equation}\label{eq:TTk}
T=(1+G)^{-1} \approx G^{-1}
\end{equation}
\begin{equation}
T_k=(1+G_k)^{-1} \approx G_k^{-1} = G^{-1} M_k
\end{equation}
From Eq.~(\ref{eq:TTk}), we have $T_k\approx TM_k$. Similarly, we have $F_k \approx F$. 
Based on the two approximations, the following learning algorithm, which transfers the learning capability among heterogeneous agents, is developed as follows.
\begin{equation}\label{eq:learningsignal2}
s = \sum_{k\in \Omega} M_k^{-1} [\alpha_{k}\{s_{k}\} + L_{k}\{h_{k}\}]
\end{equation}
Plugging it into the target UAV, we have
\begin{equation}\label{eq:error2}
\begin{split}
h&=T\{r_d\}+F\{s\}\\
&=T\{r_d\}+ F\{\sum_{k\in \Omega} M_k^{-1} \alpha_{k}\{s_{k}\} + M_k^{-1}L_{k}\{h_{k}\}\}
\end{split}
\end{equation}
Considering that for each training UAV, we have
\begin{equation}
h_k=T_k\{r_d\}+F_k\{s_k\}
\end{equation}
then
\begin{equation}\label{eq:error3}
\begin{split}
h&=T\{r_d\}+ \sum_{k\in \Omega} [M_k^{-1} \alpha_{k}F\{s_{k}\} + M_k^{-1}(FL_{k})\{h_{k}\}]\\
&=T\{r_d\}+ \sum_{k\in \Omega} [M_k^{-1} \alpha_{k}(h_k-T_k\{r_d\}) + M_k^{-1}(FL_{k})\{h_{k}\}]\\
&\approx M_k^{-1} (\sum_{k\in \Omega} \alpha_{k} I + FL_{k})\{h_{k}\}
\end{split}
\end{equation}

\begin{figure}[!htbp]
	\begin{center}
		\includegraphics[width=0.4\textwidth]{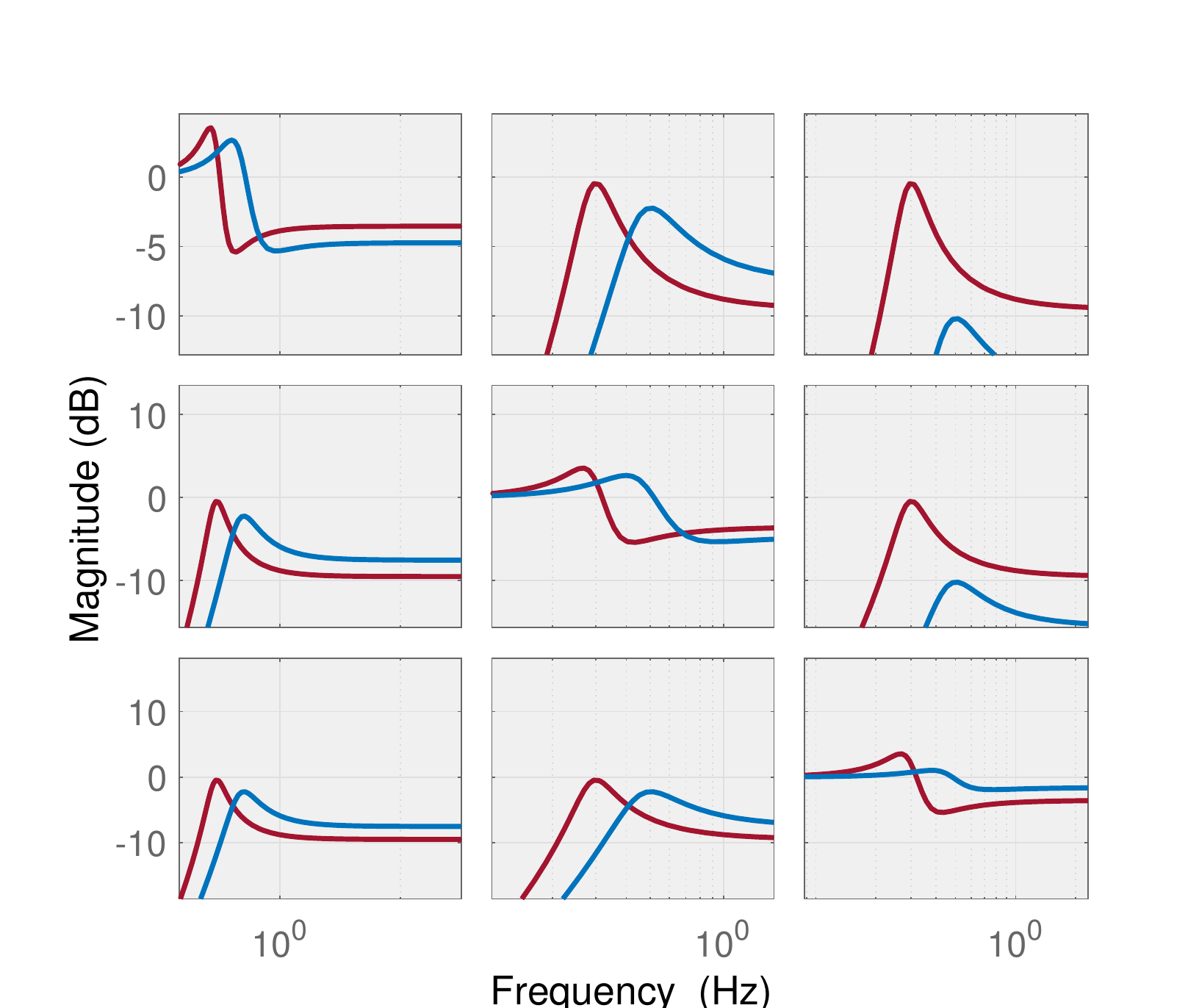} \vspace{2pt}
		\caption{Bode plots of the UAVs' closed-loop dynamics. The red and blue ones represent the training and the target UAVs, respectively.}
		\label{fig:DifferentModelInf}
	\end{center}
\end{figure}
\vspace{-25pt}
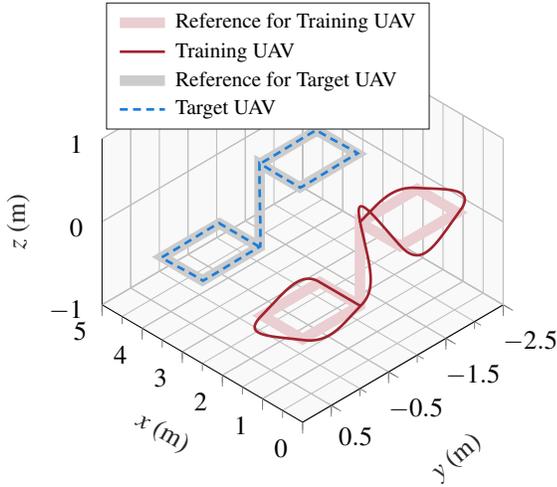
\begin{figure}[!htbp]
	\vspace{10pt}
	\begin{center}
		\input{figure/DifferentModel.tex}
		\caption{Trajectory tracking performance comparison between the training and the target UAVs. The dynamics of the two UAVs are different, as shown in Fig.~\ref{fig:DifferentModelInf}.}
		\label{fig:DifferentModel}
	\end{center}
\end{figure}

\textit{Stability and learning convergence:} 
Similarly, the learning signal is injected to the feedforward look of the UAV tracking systems and thus not affect the stability of the UAV tracking system. Based on the dynamic relationship between $h$ of the target UAV and the ones ($h_k,~k\in \Omega$) of the training UAVs in Eq.~(\ref{eq:error3}), the convergence of the learning algorithm in Eq.~(\ref{eq:learningsignal2}) for heterogeneous UAVs can be obtained as 
\begin{equation}\label{eq:opt2}
\|M_0^{-1}(\alpha_{0}I+FL_0), ..., M_N^{-1}(\alpha_{N}I+FL_N)\|_\infty{<}1/N
\end{equation}
The learning filters will be designed such that the above condition is satisfied.

\subsection{\textcolor{black}{Numerical Verification}}
This section provides a numerical study for the learning between two UAVs with different dynamics. Fig.~\ref{fig:DifferentModelInf} provides the bode plots of the dynamics from the reference $r_d\in R^{3 \times 1}$ to the real trajectory $r\in R^{3 \times 1}$ for two different UAVs. For example, the sub-figure (2,3) stands for the UAV's dynamics from the $r_{d,z}$ to $r_y$. 

Fig.~\ref{fig:DifferentModel} shows the trajectory tracking performances for both the training UAV and the target UAV. The training UAV flies one sampling time in advance, sending tracking error signal to the learning filters and generating the learning signal for the target UAV. It shows that although the tracking error of the training UAV is large while tracking an aggressive desired trajectory with sharp turns, the tracking error of the target UAV is very small. The target UAV's performance has been significantly improved in near real time via learning from another UAV with different dynamics.

\section{Unification and Verification}
In this section, we generalize the learning from UAVs with different dynamics and different flying scenarios into single unified algorithm, as detailed in \textbf{Algorithm 1}. The unified learning algorithm becomes
\begin{equation}\label{eq:unifiedlearning}
s_k{=}\Lambda_k^{-1}M_k^{-1}(\alpha_k\{s_k(t)\}{+}L_k\{h_k(t)\})
\end{equation}

\textit{Learning convergence.} Similarly, the dynamic relationship between the tracking error of the target UAV and the ones of the training UAV can be represented as follows
\begin{equation}\label{eq:error4}
\begin{split}
h\approx \sum_{k\in \Omega} \Lambda_k^{-1} M_k^{-1} (\sum_{k\in \Omega} \alpha_{k} I + FL_{k})\{h_{k}\}
\end{split}
\end{equation}
Therefore, the convergence condition for the unified learning algorithm in Eq.~(\ref{eq:unifiedlearning}) is provided as follows:
\begin{equation}\label{eq:convergence}
	\| \begin{bmatrix}
	\cdots & \Lambda_k^{-1}M_k^{-1}(\alpha_k{+}F_kL_k) & \cdots 
	\end{bmatrix}\|_\infty < 1
	\end{equation}

\begin{figure}[!htbp]
	\begin{center}
		\input{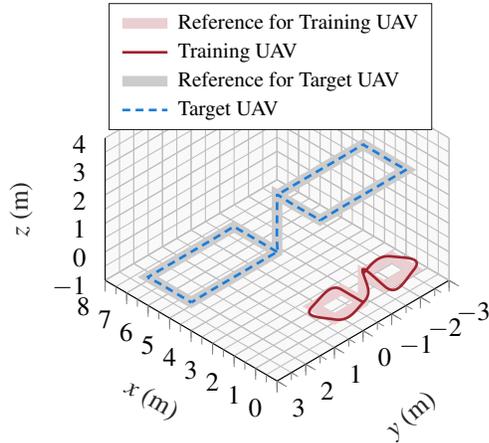}
		\caption{Trajectory tracking performance comparison between the training and the target UAVs. Both the dynamics and the references of the two UAVs are different.}
		\label{fig:DifferentModelReference}
	\end{center}
\end{figure}

Numerical verification is provided in Fig.~\ref{fig:DifferentModelReference}: the target UAV is learning from the training UAV with different dynamics flying in a transformable scenario (as shown in Fig.~\ref{fig:DifferentModelInf}). The learning process is near real time and the performance of the target UAV is significantly enhanced in terms of smaller overshoot and tracking error. 
\vspace{-15pt}
\begin{algorithm}[!htbp]
	\textbf{Inputs:}\\
	\textbf{1}. Dynamic Models $F_1$, $F_2$,..., $F_k$, $F$\\
	\textbf{2}. Dynamic Models $T_1$, $T_2$,..., $T_k$, $T$\\
	\textbf{3}. Model Scaling Factors $M_1$, $M_2$,..., $M_k$\\
	\textbf{4}. Reference Scaling factors $\Lambda_1$, $\Lambda_2$..., $\Lambda_k$\\
	\textbf{5}. Robust filters $\alpha_1$, $\alpha_2$,..., $\alpha_N$\\
	\textbf{6}. Learning filters $L_{1}$, $L_{2}$,..., $L_{N}$\\[0.5ex]
	\textbf{Initialization:} {Plan trajectory $r_d$ for the training UAV }\\[0.5ex]
	\eIf{$t<T_s$}{Target UAV is waiting.}
	{
		\For{$k=1:N$}{
			\While{Training UAV$_k$ is flying}
			{
				\While{$\|h_k(t)\|>\delta_h$}{
					1. Calculate training UAV's tracking error $h_k(t)$\\
					2. Send $h_k(t)$ into $L_k$: $L_k\{h_k(t)\}$\\
					3. Scale by $M_k$:
					$M_k^{-1}L_k\{h_k(t)\}$\\
					4. Scale by $\Lambda_k$:
					$\Lambda_k^{-1} M_k^{-1}L_k\{h_k(t)\}$
				}
				\While{$\|s_k(t)\|>\delta_s$}{
					1. Store training UAV's learning signal $s_k(t)$\\
					2. Send $s_k(t)$ into $\alpha_k$: $\alpha_k\{s_k(t)\}$\\
					3. Scale by $M_k$: {$M_k^{-1} \alpha_k\{s_k(t)\}$}\\
					4. Scale by $\Lambda_k$: \small{$\Lambda_k^{-1} M_k^{-1} \alpha_k\{s_k(t)\}$}
				}
				Calculate $s_k{=}\Lambda_k^{-1}M_k^{-1}(\alpha_k\{s_k(t)\}{+}L_k\{h_k(t)\})$
			}
		}
		Calculate $s(t)=\sum_{k=1}^{N}s_k(t)$.
		
		Send $s(t)$ into target UAV.
	}
	\caption{Physical Model Oriented Learning Algorithm}
\end{algorithm}

\vspace{-25pt}

\section{Conclusions and Future Work}
This paper has presented a physical model orientated learning algorithm and applied it to the UAV's tracking problem in presence of an aggressive desired trajectory with sharp turns and rises. This algorithm utilizes the flying data from the training UAVs flying in transformable scenarios or with different dynamics and enables the target UAV to learn from other UAVs. The learning algorithm inherits the advantages of traditional control techniques in terms of accuracy and robustness as well as the advantages of self-learning capacity and flexibility. The proposed algorithm incorporates the physical model into the learning algorithm and demonstrates great learning efficiency and reliability. As the future work, we will explore the robustness of this algorithm by explicitly including unmodeled dynamics and environmental uncertainties.
\bibliographystyle{IEEEtran}
\bibliography{bib/MZ,bib/ILC,bib/UAV2,bib/UAV}
\end{document}

%% file: figure/Reference2.tex
%
%
\definecolor{mycolor1}{rgb}{0.50196,0.50196,0.50196}%
\begin{tikzpicture}
\begin{axis}[%
scale=0.5,
width=3in,
height=3in,
at={(0.618in,0.518in)},
scale only axis,
xmin=-2.5,
xmax=6.5,
xlabel style={font=\color{white!15!black},rotate=-10},
xlabel={$x$ (m)},
tick align=outside,
ymin=-2.5,
ymax=1.5,
ylabel style={font=\color{white!15!black},rotate=60},
ylabel={$y$ (m)},
zmin=-1,
zmax=4,
zlabel style={font=\color{white!15!black}},
zlabel={$z$ (m)},
view={195}{25},
axis background/.style={fill=gray!5},
axis x line*=bottom,
axis y line*=left,
axis z line*=left,
legend style={at={(0.01,0.72)}, anchor=south west, legend cell align=left, align=left, draw=white!15!black,font=\fontsize{8}{5}\selectfont},
grid=both,
ticks=both,
xtick={-2,0,...,6},
minor xtick={-2,-1.5,...,6},
ytick={-2,-1,...,2},
minor ytick={-2.5,-2,...,2.5},
ztick={-1,0,...,4},
minor ztick={-1,0,...,4},
enlarge x limits=0,
scaled x ticks = true,
enlarge y limits=0,
scaled y ticks = true,
after end axis/.code={
\draw[black,->,>=stealth',thick] (axis cs:0.5,0,0) -- (axis cs:1,0,0);
\draw[black,->,>=stealth',thick] (axis cs:-0.5,0,1) -- (axis cs:-1,0,1);
\draw[white!60!black,->,>=stealth',thick] (axis cs:4.4,0,0) -- (axis cs:4.8,0,0);
\draw[white!60!black,->,>=stealth',thick] (axis cs:5,0.8,0) -- (axis cs:5,1.2,0);
\draw[white!60!black,->,>=stealth',thick] (axis cs:4.8,2,0) -- (axis cs:4.2,2,0);
\draw[white!60!black,->,>=stealth',thick] (axis cs:4,1.2,0) -- (axis cs:4,0.8,0);
\draw[white!60!black,->,>=stealth',thick] (axis cs:3.4,0,2) -- (axis cs:3.0,0,2);
\draw[white!60!black,->,>=stealth',thick] (axis cs:3,-0.8,2) -- (axis cs:3,-1.2,2);
\draw[white!60!black,->,>=stealth',thick] (axis cs:3.4,-2,2) -- (axis cs:3.8,-2,2);
\draw[white!60!black,->,>=stealth',thick] (axis cs:4,-1.2,2) -- (axis cs:4,-0.8,2);
}
]
\addplot3 [color=white!10!black, line width=2.0pt]
 table[row sep=crcr] {%
	0	0	0\\
2	0	0\\
2	1	0\\
0	1	0\\
0	0	0\\
0	0	1\\
-2	0	1\\
-2	-1	1\\
0	-1	1\\
0	0	1\\
};
 \addlegendentry{Test reference for Training UAV}
 
 \addplot3 [color=white!60!black, line width=2.0pt]
 table[row sep=crcr] {%
 		4	0	0\\
 	5	0	0\\
 	5	2	0\\
 	4	2	0\\
 	4	0	0\\
 	4	0	2\\
 	3	0	2\\
 	3	-2	2\\
 	4	-2	2\\
 	4	-0	2\\
 };
 \addlegendentry{Test reference for Target UAV}
 
\addplot3[only marks, mark=*, mark options={fill=red}, mark size=2.5000pt, draw=red,line width=1.0pt] table[row sep=crcr]{%
	x	y	z\\
	0	0	0\\
4 0	0\\
};\addlegendentry{Start point}

\addplot3[only marks, mark=*, mark options={fill=blue}, mark size=2.5000pt, draw=blue,line width=1.0pt] table[row sep=crcr]{%
	x	y	z\\
	0	0	1\\
	4 0	2\\
};\addlegendentry{End point}

\end{axis}
\end{tikzpicture}%

%% file: figure/DifferentModel.tex
%
\definecolor{mycolor1}{rgb}{0.00000,0.44700,0.74100}%
\definecolor{mycolor2}{rgb}{0.85000,0.32500,0.09800}%
\definecolor{mycolor3}{rgb}{0.92900,0.69400,0.12500}%
\definecolor{mycolor4}{rgb}{0.49400,0.18400,0.55600}%
\definecolor{mycolor1}{rgb}{0.50196,0.50196,0.50196}%
\begin{tikzpicture}

\begin{axis}[%
scale=0.7,
width=3in,
height=3in,
at={(0.618in,0.518in)},
scale only axis,
xmin=0,
xmax=5,
xlabel style={font=\color{white!15!black},rotate=-30},
xlabel={$x$ (m)},
tick align=outside,
ymin=-2.5,
ymax=1,
ylabel style={font=\color{white!15!black},rotate=40},
ylabel={$y$ (m)},
zmin=-1,
zmax=1,
zlabel style={font=\color{white!15!black}},
zlabel={$z$ (m)},
view={225}{45},
axis background/.style={fill=gray!5},
axis x line*=bottom,
axis y line*=left,
axis z line*=left,
legend style={at={(0.01,0.73)}, anchor=south west, legend cell align=left, align=left, draw=white!15!black,font=\fontsize{8}{5}\selectfont},
grid=both,
ticks=both,
xtick={-1,...,5},
minor xtick={-1,-0.5,...,5},
ytick={-2.5,-1.5,...,0.5},
minor ytick={-2.5,-2,...,0.5},
ztick={-1,0,...,2},
minor ztick={-1,0,...,2},
enlarge x limits=0,
scaled x ticks = true,
enlarge y limits=0,
scaled y ticks = true,
]
\addplot3 [color=white!80!red, line width=4.0pt]
table[row sep=crcr] {%
0	0	0\\
1	0	0\\
1	1	0\\
0	1	0\\
0	0	0\\
0	0	1\\
-1	0	1\\
-1	-1	1\\
0	-1	1\\
0	0	1\\
};
 \addlegendentry{Reference for Training UAV}

\addplot3 [color=white!10!red, line width=1.0pt]
table[row sep=crcr] {%
0	0	0\\
1.05159010877622	0.00125286615125009	0\\
1.09947446320499	0.00604204623349269	0\\
1.14218798335902	0.0146460888340054	0\\
1.17945950204404	0.0272632422843533	0\\
1.2111228438701	0.0440144607385382	0\\
1.23711145111518	0.0649473820325259	0\\
1.25745187075736	0.090041141984422	0\\
1.2722562900106	0.119211887066106	0\\
1.28171430827525	0.152318846997949	0\\
1.28608413102975	0.189170830616301	0\\
1.28487988415763	0.240140914349539	0\\
1.27695278017635	0.296055665748437	0\\
1.25972638527278	0.368824092509911	0\\
1.23553992964392	0.446705707484294	0\\
1.19523618042045	0.55645894333779	0\\
1.12515174788421	0.728133747255333	0\\
1.02603439668398	0.971129330276795	0\\
0.988478207194798	1.06402706346073	0\\
0.958419290808342	1.12149704063992	0\\
0.932022356108498	1.16151678262894	0\\
0.903477340884768	1.19599861849292	0\\
0.872769194364429	1.22482692898253	0\\
0.84832913097335	1.24272309398647	0\\
0.822702879101073	1.25745187075736	0\\
0.795919911277222	1.26906505839833	0\\
0.768018211178659	1.27763896303296	0\\
0.739043450368706	1.28327231892207	0\\
0.709048144459244	1.28608413102975	0\\
0.667568761626885	1.28568336610763	0\\
0.624532138660652	1.28087959530576	0\\
0.580103357091034	1.27208088465605	0\\
0.522875601336413	1.25613584580713	0\\
0.452186568483095	1.23094432341778	0\\
0.367712796802186	1.19523618042045	0\\
0.245083904856859	1.13721320779872	0\\
0.0017371932698349	1.02070302464169	0\\
-0.0658506078122643	0.988478207194798	0\\
-0.114743755609488	0.958419290808342	0\\
-0.148592857952986	0.932022356108498	0\\
-0.177572027153952	0.903477340884768	0\\
-0.20160043919297	0.872769194364429	0\\
-0.220680220387033	0.839917574285743	0\\
-0.234890131069078	0.80497391273952	0\\
-0.24437853847512	0.768018211178659	0\\
-0.249925228248359	0.719156476093387	0\\
-0.248954767035639	0.667568761626885	0\\
-0.242101560652202	0.613548783094172	0\\
-0.227124997322359	0.545972702083208	0\\
-0.20605225877006	0.475959697350108	0\\
-0.170893419007913	0.379878862222613	0\\
-0.114967481739162	0.245083904856859	0\\
-0.0183924878345054	0.0136649375205167	0\\
0.00795437223989581	-0.0552344135273293	0.00125286615125009\\
0.0237670547694717	-0.0960422783824968	0.00604204623349269\\
0.0380131301364193	-0.132268601189129	0.0146460888340054\\
0.0505901887154578	-0.163698248284193	0.0272632422843533\\
0.0614294787606688	-0.190206818924676	0.0440144607385382\\
0.070494423808477	-0.211755702955704	0.0649473820325259\\
0.0777787530739054	-0.22838618151356	0.090041141984422\\
0.0833043031684431	-0.240212734799715	0.119211887066106\\
0.0871185501501961	-0.24741572084167	0.152318846997949\\
0.0895891027397189	-0.250285698181957	0.198942327710748\\
0.0896782712385988	-0.246880872759823	0.250946821545884\\
0.08695553771159	-0.235461158545582	0.31967223188025\\
0.0815633005298015	-0.217263879470707	0.394274287744347\\
0.071057343933892	-0.184844257180777	0.500884730666653\\
0.0524531568832822	-0.130694189500484	0.656031704583549\\
0.0134007789126145	-0.0230552100842631	0.957146860099644\\
-0.00197273967891687	0.0120475141881116	1.06402706346073\\
-0.0170627376383528	0.0310931222829935	1.12149704063992\\
-0.0374163840312045	0.0476067169864873	1.17066264582557\\
-0.0576406876623117	0.0588851058769937	1.20373805976378\\
-0.0814169724047238	0.0683953736275427	1.23114637786439\\
-0.101571833490463	0.0743584316441526	1.24798258672778\\
-0.123688766300716	0.0793235898317319	1.26166582635738\\
-0.147724739614066	0.0833043031684431	1.2722562900106\\
-0.17362226540475	0.0863219372737232	1.27983799521549\\
-0.201310735208461	0.0884051261628167	1.28451667604235\\
-0.230707795543646	0.0895891027397189	1.28641760475534\\
-0.27240015015538	0.0898408376108537	1.28487988415763\\
-0.31672038665102	0.0886793135205914	1.27903774861008\\
-0.375409102155093	0.0854316024951915	1.26631970021953\\
-0.449922703568062	0.0792551446292906	1.24427716377465\\
-0.554565080480174	0.0679628934632261	1.20599597228757\\
-0.704298508960132	0.0487627738240224	1.14321693527712\\
-0.977200928642376	0.0116554840267922	1.02603439668398\\
-1.05392898691095	0.000444098955412642	0.995365171107239\\
-1.09950620713593	-0.0104126489315086	0.976869114187129\\
-1.14008686130562	-0.0245568281408128	0.960133052356859\\
-1.16708837858284	-0.0374163840312045	0.948813515010268\\
-1.19106723945294	-0.0522521837805467	0.93859479122364\\
-1.21198535726785	-0.0690846005704133	0.929503857075127\\
-1.22983382511976	-0.0879152292770875	0.921556194348691\\
-1.24463132045302	-0.108727949653977	0.914756313378736\\
-1.25642235791235	-0.131490084764872	0.909098342665779\\
-1.26527541454304	-0.156153638986799	0.904566676232502\\
-1.27128095120091	-0.18265660009899	0.901136669698318\\
-1.27454935357681	-0.210924290297555	0.898775376138477\\
-1.27520881560656	-0.24087075141359	0.897442312955333\\
-1.27228036069133	-0.283243903801916	0.897181863104114\\
-1.26534836126476	-0.328180275245258	0.898536719926565\\
-1.25482365176255	-0.375409102155093	0.901363501870856\\
-1.23728056805809	-0.437232477963967	0.906729317542038\\
-1.21568889537937	-0.501612868530251	0.913830034412941\\
-1.18569840298124	-0.581421523724306	0.924171740139959\\
-1.14128978488904	-0.690540048411251	0.940102524490304\\
-0.964275717356333	-1.11013137347094	1.00511166569785\\
-0.939207210028699	-1.14941962181213	1.01144222222457\\
-0.911933683144407	-1.18341310240506	1.01711568610786\\
-0.882424667421243	-1.21198535726785	1.0220937874456\\
-0.858828483141556	-1.22983382511976	1.02535574815948\\
-0.833993295714051	-1.24463132045302	1.02820695096787\\
-0.807944630518351	-1.25642235791235	1.03064581292776\\
-0.780717251138355	-1.26527541454304	1.03267447862698\\
-0.752354322792883	-1.27128095120091	1.03429858743358\\
-0.722906549657605	-1.27454935357681	1.03552702327438\\
-0.692431295184603	-1.27520881560656	1.03637165015193\\
-0.650309160432107	-1.27228036069133	1.03692631746167\\
-0.606635945985984	-1.26534836126476	1.03686593061328\\
-0.56158614564864	-1.25482365176255	1.03623801584859\\
-0.503616804385937	-1.23728056805809	1.03473576964832\\
-0.444148391633776	-1.21568889537937	1.03254089499095\\
-0.371324914223107	-1.18569840298124	1.02915860925697\\
-0.272759191055052	-1.14128978488904	1.02372027221078\\
0.0857260049013837	-0.976003356722373	1.00128993434609\\
0.132122602446397	-0.945679016479158	0.998146182493271\\
0.170998225464877	-0.911933683144407	0.995341706092401\\
0.202125290188842	-0.874698110690132	0.992911985243254\\
0.225483482631502	-0.833993295714051	0.990880180668959\\
0.241240810147764	-0.789921511484957	0.989257733166707\\
0.24973191097898	-0.742656232475245	0.988045188070637\\
0.251434399372441	-0.692431295184603	0.987233203844069\\
0.246943998087741	-0.639529626957912	0.986803703333604\\
0.234355288941381	-0.572967768761156	0.986756743661847\\
0.211411992575047	-0.491831242903111	0.987276365949072\\
0.177021864123735	-0.395741648490476	0.98849895920931\\
0.116096455088577	-0.248000253679131	0.991226120781105\\
0.0236437901334561	-0.0270050168033249	0.996174921574243\\
0.0143277686080345	-0.0029409747142668	0.996722773132264\\
};
 \addlegendentry{Training UAV}

\addplot3 [color=white!80!black, line width=4.0pt]
table[row sep=crcr] {%
2.5	0	0\\
3.5	0	0\\
3.5	1	0\\
2.5	1	0\\
2.5	0	0\\
2.5	0	1\\
1.5	0	1\\
1.5	-1	1\\
2.5	-1	1\\
2.5	0	1\\
};
 \addlegendentry{Reference for Target UAV}

\addplot3 [color=white!10!blue, line width=1.0pt, densely dashed]
table[row sep=crcr] {%
2.5	0	0\\
3.51325092645465	0	0\\
3.51550860492625	0.12641333528362	0\\
3.51215383867405	0.278398508146801	0\\
3.49755248452452	0.751719847933876	0\\
3.49696608096013	1.00278388077066	0\\
3.49708396691127	1.01325092645465	0\\
3.40109245484324	1.01554959336278	0\\
3.27176163426536	1.01373061707816	0\\
3.07723319504082	1.00665265223396	0\\
2.85087370014051	0.999343909217704	0\\
2.66798082422722	0.996842899198799	0\\
2.4869641776686	0.997083966911272	0\\
2.48450475743764	0.871693277442864	0\\
2.48772899213101	0.720953551668122	0\\
2.50252746754116	0.238821160927858	0\\
2.50293464982137	-0.0130358223313989	0\\
2.50209709155779	-0.0155107473679736	0.112691610925899\\
2.50099275069766	-0.0135545472711427	0.246809558423128\\
2.49946935466452	-0.00332921867892066	0.520322226667841\\
2.49935623565613	0.00263158980802114	0.770271958379308\\
2.49974264203113	0.0030491317551018	1.00001592525764\\
2.49976396356285	0.00298449796742251	1.01002862360759\\
2.05646104772168	-0.000282561372975998	0.99992952277405\\
1.74833982001283	-0.000665703187463507	0.999732672959109\\
1.48673107695071	-0.000221681834625009	0.99998394799789\\
1.48450034538621	-0.13642663374579	1.00001316429926\\
1.49008541777252	-0.340067332982632	1.00000813098718\\
1.50109772042027	-0.670195681794266	0.999996973099789\\
1.50333273222237	-0.892858050096626	0.999997034174892\\
1.50292265647723	-1.01326892304929	1.00000161439992\\
1.59891307887681	-1.01557901034027	1.00000163595335\\
1.72824202838236	-1.01376319108561	0.999998417963211\\
1.9124214471685	-1.00705879274923	0.999998722686434\\
2.20016208450083	-0.998302085530764	0.999999017767056\\
2.43254779338184	-0.996698010633233	1.00000078515899\\
2.5130351754416	-0.997077343522772	0.99999928020846\\
2.51549492530498	-0.871687824109598	1.00000062917563\\
2.51227120940659	-0.720951134591457	0.999999457506973\\
2.49747267136607	-0.238822549349949	1.00000033804764\\
2.49695083869602	0.00253248526216288	1.0000002655658\\
};
 \addlegendentry{Target UAV}
\end{axis}
\end{tikzpicture}%